\documentstyle[aps,prd,multicol]{revtex} 

\begin{document}

\draft



\preprint{UNIL-IPT-01-03}

\title%
{On the thin-string limit of the 6d stringlike defect model}

\author%
{P. Tinyakov\footnote{Email: Peter.Tinyakov@ipt.unil.ch} and K. Zuleta\footnote{Email: Katarzyna.ZuletaEstrugo@ipt.unil.ch}}

\address%
{%
Institute of Theoretical Physics,
University of Lausanne, \\
CH-1015 Lausanne,  Switzerland
}

\date{\today}

\maketitle

\begin{abstract}%
We show that in 6d models localizing gravity on stringlike defects and satisfying the dominant energy condition, the metric exterior to the string inevitably depends on the string's thickness. As a consequence, in the limit of thin string either the gravity delocalizes, or the six-dimensional Planck scale must be much larger than the four-dimensional one.  
\end{abstract}

\begin{multicols}{2}

Since the work by  Randall and Sundrum~\cite{rs}, it is now quite generally agreed that the scenarios modeling our four-dimensional world as a topological defect in a higher dimensional spacetime can be perfectly compatible with  four-dimensional gravity, without the need of compactifying the extra dimensions.
Studying a thin 3-brane embedded in a  five-dimensional anti-de Sitter spacetime and fine-tuning   the bulk cosmological constant to the tension of the brane, the authors of~\cite{rs} constructed a solution to the Einstein equations preserving  four-dimensional Poincar\'e invariance. Due to the presence of a single graviton zero mode localized on the brane and  to the strong suppression of higher-dimensional corrections,   gravity had  an essentially four-dimensional character. The metric solution and the shape of the zero mode were entirely determined by the brane tension and bulk cosmological constant.

Another interesting proposal, that gravity can be localized on a local stringlike  defect in a 6d  spacetime, was made in~\cite{shap1} and subsequently generalized to higher dimensions~\cite{shap2}. Clearly, a very appealing feature of this idea is the fact that the defect itself arises naturally from a simple field-theoretical model~(the abelian Higgs model). Moreover, assuming the existence of a thin-string limit, it  would produce an effectively four-dimensional gravity without resorting  to fine-tuning of the cosmological constant in the bulk to the tension on the brane. The  necessary tuning would involve some  relations between the tension components. 

The aim of this work is to demonstrate  an essential difference between the 5d and the  6d models localizing gravity. This difference  lies  in the dependence of the metric exterior to the defect on the defect's intrinsic characteristics and in particular on its transverse dimensions.  In the 5d scenario proposed in~\cite{rs}, the dependence of the exterior part of the metric on the brane's thickness disappears in the thin-brane limit and it then suffices to characterize the defect by one parameter: its tension. An unambiguous exact exterior metric, without any dependence on the  brane structure, can be then  determined. In the six-dimensional case the situation is quite distinct. 
As we will show here, requiring physical fields as  source of the defect, unavoidably yields the dependence of the exterior metric 
on the string's width even \emph{in} the thin-string limit.

Let us remind the model introduced in \cite{shap1} . Its setup is a~6d spacetime equipped with the metric:
\begin{equation}\label{metric}
    ds^2 = \sigma(\rho) g_{\mu\nu} dx^\mu dx^\nu -d\rho^2 
    - \gamma(\rho) d\theta^2\, ,
\end{equation}
with a negative cosmological constant in the bulk and a singular stringlike defect (3-brane) localized at~$\rho=0$. The defect is  modeled by the continuous matter distribution within a core of radius~$\epsilon$ and  vanishing for~$\rho>\epsilon$, parametrized by the stress-energy tensor~$T^A_B$:
\begin{eqnarray}
\label{source}
    & T^\mu_\nu  = \delta^\mu_\nu f_0(\rho), \quad T_\rho^\rho = 
     f_\rho(\rho), \quad  T_\theta^\theta = f_\theta(\rho)~,\quad \nonumber\\
     & {\rm and}  \quad \quad T^A_B=0 \quad {\rm otherwise} \ .
\end{eqnarray}
Within the classical theory of relativity, one expects to be able to take a thin-string limit reducing the string's transverse dimensions to zero.

After setting the four-dimensional cosmological constant~$\Lambda_{phys}$ to zero, the Einstein equations become:
\begin{eqnarray}
\label{solnset}
    \frac{3}{2} \frac{\sigma^{\prime\prime}} {\sigma} 
    + \frac{3}{4}\frac{\sigma^\prime}{\sigma}\frac{\gamma^\prime}{\gamma}
    -\frac{1}{4}\frac{\gamma^{\prime 2}}{\gamma^2}
    +\frac{1}{2}\frac{\gamma^{\prime\prime}}{\gamma}
    &=& -\frac{1}{M_6^4}
    (\Lambda +  f_0(\rho)) 
        ~, \nonumber\\
    \frac{3}{2}\frac{\sigma^{\prime 2}}{\sigma^2}+
    \frac{\sigma^\prime}{\sigma}
    \frac{\gamma^\prime}{\gamma}&=
    & -\frac{1}{M_6^4}
    (\Lambda +  f_\rho(\rho)) 
        ~,\\
    2\frac{\sigma^{\prime\prime}}{\sigma} + 
    \frac{1}{2}\frac{\sigma^{\prime 2}}{\sigma^2}
    &=& -\frac{1}{M_6^4}
    (\Lambda +  f_\theta(\rho))
        ~, \nonumber
\end{eqnarray}
$M_6$ being the six-dimensional Planck scale, related to the four-dimensional Planck scale~$M_P$ by:
\begin{equation}\label{Planck}
M_P^2=2\pi M_6^4\int_0^\infty d\rho\, \sigma\sqrt\gamma\ .
\end{equation}
In order to guarantee the regularity of the geometry, the solution inside the core has to satisfy the boundary conditions at the origin:
\begin{eqnarray}
\label{bconds}
& \sigma\big|_{\rho=0} =A~, \quad   \sigma^\prime\big|_{\rho=0} =0~, \nonumber\\
& \quad (\sqrt\gamma)^\prime\big|_{\rho=0} = 1~, 
\quad {\rm and}  \quad \gamma\big|_{\rho=0} = 0\ ,
\end{eqnarray}
 where~$A$ is an arbitrary constant.  
 The metric outside the core is  the solution of~Eqns.~$(\ref{solnset})$ in the absence of sources:
\begin{equation}
\label{solnform}
     \sigma(\rho) =  e^{-c \rho}~ \quad \mbox{and} \quad \gamma(\rho)=R_0^2e^{-c\rho}\ ,
\end{equation}
where~$c=\sqrt{\frac{2(-\Lambda)}{5 M_6^4}}$ and~$R_0$ is a constant.
The exterior solution has to be adjusted to the solution inside the string core by means of the junction conditions at~\mbox{$\rho=\epsilon$}, translating into the relations between the components of the string tension per unit length,~$\mu_i$, defined by~:
$$
\mu_i=\int_0^\epsilon d\rho\, \sigma\sqrt\gamma f_i \ ,
$$
where~$i=0,\rho, \theta$.
 In the limit~$\epsilon\to 0$ these relations  can be written:
\begin{equation}\label{tensions}
\mu_0=\mu_\theta+A^2M_6^4\ , \qquad\qquad \mu_\rho+\mu_\theta= 2 c R_0M_6^4\ .
\end{equation}
 A priori, as it was assumed in~\cite{shap1}, it could seem that as far as  the constant~$R_0$  is  such that  the  Eqns.~(\ref{tensions}) are satisfied, it can be otherwise arbitrary. This would imply a solution with a finite volume transverse space in the thin-string limit. The shape of the localized graviton zero mode:
\begin{equation}\label{zero-mode}
\psi_0(\rho)=\sqrt\frac{3c}{2R_0}\,e^{-{3}/{4}c\rho} \ , 
\end{equation}
would then depend on the cosmological constant and string's tension components, like in~\cite{rs}.

We will now  prove that (under some assumptions on the regularity of the metric components) this impression is not correct and that in fact, in order to get a physical solution, the exterior part of the metric must depend on the string's width. To see that it must  indeed be the case, let us first remark that the equation involving $ f_0(\rho)$ can be rewritten in the  form:
\begin{equation}
\label{fzero}
\frac{1}{M_6^4} f_0(\rho)=-{\frac{1}{M_6^4}\Lambda}-\frac{3}{4} \frac{\sigma^{\prime\prime}} {\sigma}- \frac{3}{16} \frac{{\sigma^{\prime}}^2} {\sigma^2}-\frac{{\left(\sigma^{3/4}\gamma^{1/2}\right)}^{\prime\prime}} {\sigma^{3/4}\gamma^{1/2}} \ ,
\end{equation}
 which can be further rewritten as:
\begin{equation}
\label{fzero_prime}
 \frac{1}{M_6^4}f_0(\rho)=-\frac{5}{8M_6^4}{\Lambda}+ \frac{3}{8M_6^4}  f_\theta(\rho) -\frac{{\left(\sigma^{3/4}\gamma^{1/2}\right)}^{\prime\prime}} {\sigma^{3/4}\gamma^{1/2}} \ .
\end{equation}

By means of a simple argument (which we postpone to the Appendix),  it can be proved that if we decrease~$\epsilon$, while keeping $R_0 \sim{\cal{O}}(1)$, at some point the last term in the  equation~(\ref{fzero_prime}) must become  hugely negative somewhere inside  the interval~$(0,\epsilon)$. More precisely, for a sufficiently small~$\epsilon$ and  some $\rho^\#<\epsilon$ we have:    
\begin{equation}  
\label{estim}
\left.\frac{{\left(\sigma^{3/4}\gamma^{1/2}\right)}^{\prime\prime}} {\sigma^{3/4}\gamma^{1/2}}\right|_{\rho=\rho^\#}
\!\!\sim {\cal{O}}\left(\frac{1}{\epsilon^2}\right) \ .
\end{equation}

Consequently, inserting this result into (\ref{fzero_prime}), for sufficiently small~$\epsilon$ and suitably chosen po\-si\-tive constant~$C$~we get:
\begin{equation}
\label{cute_ineq}
\left.\frac{1}{M_6^4} f_0(\rho)\right|_{\rho=\rho^\#}\leq \left.\frac{3}{8M_6^4} f_\theta(\rho)\right|_{\rho=\rho^\#}\!-\frac{C}{\epsilon^2}\ .
\end{equation}
Now this inequality means violation of the dominant energy condition \cite{hawking} requiring  the energy density to  be not only positive but also greater or equal than  any of the pressure components. As a consequence, no classical configu\-ra\-tion of fields will produce an infinitely thin string allowing for a metric solution (\ref{solnform}) with constant~$R_0$.

Now that we have seen that there is no thin-string limit for a constant~$R_0$, 
let us check whether we can reach it by renouncing to the idea of arbitrary~$ R_0$ and  allowing it to change with~$\epsilon$. If we take~$ R_0\sim{\cal{O}}(\epsilon^\kappa)$, with $\kappa<1$, the inequality~(\ref{estim}) will still be valid, invariably leading to the violation of the dominant energy condition. Let us now suppose that~$ R_0\sim{\cal{O}}(\epsilon^\kappa)$, with $\kappa\geq 1$. Then the argument used to show~(\ref{estim}) is no longer valid and we can rightly hope  to find a solution with physical~$f$'s with characteristic feature $\mu_0=A^2M_6^4$, $\mu_\rho=\mu_\theta=0$.  

Assume~$M_6$ is fixed. In the limit~$\epsilon\to 0$ the four-dimensional  Planck scale~(\ref{Planck}) becomes:
$$
M_P^2= 2\pi c^{-1} R_0 M_6^4 \ .
$$
We see that letting $R_0$ go as~$\epsilon^\kappa$ will imply the Planck mass to vary in the same way. The only way to make~$M_P^2$ remain constant is to further impose~$c\sim{\cal{O}}(\epsilon^\kappa)$. But, as can be seen from the form of the zero-mode wave function~(\ref{zero-mode}), it is precisely~$c$ that regulates the localization of the gravity on the brane. As a consequence, allowing~$c\sim{\cal{O}}(\epsilon^\kappa)$ would spoil the localization of the gravity on the brane. 

Alternatively, one could fix~$M_P$ and force the  localization of gravity on the brane (which means fixing~$c$). By Eqn.~(\ref{Planck}) and given~$R_0\sim{\cal{O}}(\epsilon^\kappa)$ this would imply~$M_6^4\sim{\cal{O}}(\epsilon^{-\kappa})\to\infty$, that is making the string thin would require~$M_6\gg M_P$.
 
In conclusion, we have shown that in a 6d spacetime with a local stringlike defect, requiring the dominant energy condition to be satisfied implies that the metric exterior to the defect must depend on the defect's width.
In fact, for finite values of parameters such a metric solution can indeed be found~\cite{harvey}.
As a consequence
of the dependence of the exterior metric on the string's width, in  the thin-string limit either the  solution  becomes degenerate and  it does not lead to the localization of gravity on the defect, or a scale much higher than the four-dimensional Planck mass~$M_P$ must be introduced.

\noindent {\it Acknowledgments:}
We wish to thank S.~Dubovsky, T.~Gherghetta, H.~Meyer, E.~Roessl and M.~Shaposhnikov  for helpful discussions. This work was partly supported by  Swiss National Science Foundation, grant no.~\mbox{21-58947.99}.

\appendix
\section*{Proof of Eqn. (\lowercase{\ref{estim}})}

It remains to be shown that taking constant~$R_0$ implies~(\ref{estim}). The proof is straightforward.
For the sake of clarity of notation, let us denote~$\alpha=\sigma^{3/4}\gamma^{1/2}$. In terms of~$\alpha$, the boundary conditions at~$\rho=0$ and the junction condition read:
\begin{equation}
\label{bconds_alpha}
   \alpha\big|_{\rho=0} =0~,\quad \alpha^\prime\big|_{\rho=0} = A^{3/4}~,\quad\! {\rm and}  \quad\! \alpha\big|_{\rho=\epsilon} = R_0e^{-5/4c\epsilon}
\end{equation}

Assuming that~$\alpha$ is a continuously differentiable function, it is clear that there must exist a~$\rho^\star<\epsilon$ such that:
\begin{equation}
\alpha^\prime(\rho^\star)=\frac{\alpha(\epsilon)-\alpha(0)}{\epsilon}=\frac{R_0e^{-5/4c\epsilon}}{\epsilon}\ .
\end{equation}
This implies, assuming  further that~$\alpha^\prime$ is conti\-nu\-ously  differentiable
in~$(0,\rho^\star)$, that there exists a~\mbox{$\rho^{\star\star}<\rho^{\star}$} such that:
\begin{eqnarray}
\label{ineq}
\alpha^{\prime\prime}(\rho^{\star\star}) & = &\frac{\alpha^\prime(\rho^\star)-\alpha^\prime(0)}{\rho^\star}=\frac{R_0e^{-5/4c\epsilon}-\epsilon A^{3/4}}{\rho^\star\epsilon}\nonumber \\ 
& \geq &\frac{ R_0e^{-5/4c\epsilon}-\epsilon A^{3/4}}{\epsilon^2} \ .
\end{eqnarray}
Suppose first that 
\begin{equation}
\label{cond}
\alpha(\rho^{\star\star})< R_0e^{-5/4c\epsilon} \ .
\end{equation} 
The equation (\ref{ineq}) implies:
\begin{equation}  
\frac{\alpha^{\prime\prime}(\rho^{\star\star})}{\alpha(\rho^{\star\star})}>\frac{1}{\epsilon^2}\left[1-{\epsilon}\frac{A^{3/4}}{R_0}e^{5/4c\epsilon}\right] \ .
\end{equation}
For $\epsilon\to 0$ and $A,R_0\sim{\cal{O}}(1)$, the first term in this inequality becomes dominant and therefore: 
$$
\frac{\alpha^{\prime\prime}(\rho^{\star\star})}{\alpha(\rho^{\star\star})}\sim {\cal{O}}\left(\frac{1}{\epsilon^2}\right) \ .
$$
Suppose now that 
$$\alpha(\rho^{\star\star})\geq R_0 e^{-5/4c\epsilon} \ . $$
Then we can repeat the same reasoning to find an another point, now inside the interval $(0,\rho^{\star\star})$, satisfying an inequality analogous to~(\ref{ineq}). If this new point \mbox{still} does not satisfy~(\ref{cond}), we can repeat this procedure over and over until we find a point satisfying both~(\ref{ineq}) and~(\ref{cond}). Given that $\alpha$ is continuously differentiable and that $\alpha(0)=0$, this iteration must eventually stop and such a point must exist. 

In conclusion, there  exists a point~$\rho^\#\in (0,\epsilon)$ where, for~$\epsilon$ sufficiently small:    
$$
\frac{\alpha^{\prime\prime}(\rho^\#)}{\alpha(\rho^\#)}\sim {\cal{O}}\left(\frac{1}{\epsilon^2}\right) \ .
$$
(In fact, given the regularity properties of~$\alpha$, it will be also valid for the points in a small  neighborhood of~$\rho^\#$.)

\end{multicols}
\end{document}